\documentclass[aps,prl,twocolumn,showpacs,groupedaddress]{revtex4}
\usepackage{graphics}
\usepackage{graphicx}

\begin{document}


\title{Interference-induced enhancement of intensity and energy of a multimode quantum optical field by a subwavelength array of coherent light sources

}


\author{S.V. Kukhlevsky}
\affiliation{Department of Physics, University of P\'ecs,
Ifj\'us\'ag u.\ 6, H-7624 P\'ecs, Hungary}


\begin{abstract}
Recently, we have showed a mechanism that could provide a great transmission enhancement of the light waves 
passed through subwavelength aperture arrays in thin metal films not by the plasmon-polariton waves, but by the constructive interference of diffracted waves (beams generated by the apertures) at the detector placed in the far-field zone. We now present a quantum reformulation of the model. The Hamiltonian describing the interference-induced enhancement of the intensity and energy of a multimode quantum optical field is derived. Such a field can be produced, for instance, by a subwavelength array of coherent light sources.
\end{abstract}


\maketitle

%
\maketitle
%
%
Since the demonstration of enhanced transmission of light through a subwavelength metal apertures in the study~\cite{Ebb}, the phenomenon attracts increasing interest of researchers working in the field of nanooptics and nanophotonics~\cite{Barn,Schr,Sobn,Port,Asti,Taka,Lala,Barb,Kukh1,Leze,More,Liu,Garc,Tre,Cao,Sar,
Bar,Koer,Miro,Pend,Gome,Mo,Ung,Hua,Ben,Sary}. It is generally accepted~\cite{Liu} that the excitation and interference of plasmon-polaritons play a key role in the process of enhancement in the most of experiments (see, the recent reviews \cite{Garc,Sary}). Recently, we have showed a mechanism that could provide a great transmission enhancement of the light waves passed through subwavelength aperture arrays in thin metal films not by the plasmon-polariton waves, but by the constructive interference of diffracted waves (beams generated by the apertures) at the detector placed in the far-field zone~\cite{Kukh}. According to the model, the beams generated by multiple, subwavelength apertures can have similar phases and can add coherently. If the spacing of the apertures is smaller than the optical wavelength, then the phases of the multiple beams are nearly the same and beams add coherently (the light power and energy scales as the number of light-sources squared, regardless of periodicity). If the spacing is larger, then the addition is not so efficient, but still leads to enhancements and resonances (versus wavelength) in the total power transmitted. The analysis~\cite{Kukh} is based on calculation of the energy flux (intensity) of a beam array by using Maxwell's equations for classic, non-quantum electromagnetic fields. The enhancement mechanism was interpreted as a non-quantum analog of the superradiance emission of a subwavelength ensemble of atoms (the light power and energy scales as the number of light-sources squared, regardless of periodicity) predicted by the Dicke quantum model~\cite{Dick}. We now present a quantum reformulation of our model. The Hamiltonian describing the interference-induced enhancement of the intensity and energy of a multimode quantum optical field is derived. Such a field can be produced, for instance, by a subwavelength array of coherent light sources. 

With the objective of quantizing the electromagnetic (EM) field, it is convenient to begin with consideration of the Hamiltonian of a classical (non-quantum) EM field based on Maxwell's equations. The detailed description of the problem can be found in textbooks (for example, see~\cite{Land,Loud,Bere,Cohe,Wein,Pike,Scul}. Here, we use the theoretical formulation and notations of the studies~\cite{Land,Bere}. Let us first consider a single-mode EM field in free space, which is a superposition of $N$ time-harmonic phase-coherent plane waves with different phases. 
The vector potential of the
$n$-th linearly polarized wave is assumed to be
${\mathbf{A}_n(\mathbf{r},t)}=\mathbf{a}_{\mathbf{k}}e^{i{\mathbf{k}
{\mathbf{r}+i\varphi_{n}}}}+\mathbf{a^*}_{\mathbf{k}}e^{-i{\mathbf{k}
{\mathbf{r}-i\varphi_{n}}}}$, where
$\mathbf{k}\equiv{(k_x,k_y,k_z)}$ is the wave vector, and
$\varphi_{n}$ is the wave phase. The vector potential
$\mathbf{A}(\mathbf{r},t)=\sum_{n=1}^N{\mathbf{A}_n(\mathbf{r},t)}$
determines the electric
$\mathbf{E}(\mathbf{r},t)=-{\frac{1}{c}}\mathbf{\dot{A}}$ and
magnetic $\mathbf{H}(\mathbf{r},t)=\mathbf{\nabla} \times
\mathbf{A}$ components of the field. One can easily find the
Hamiltonian ${\cal H}$ of the field by calculating the field
energy ${\cal E}={\frac{1}{8\pi
}}{\int}(\mathbf{E}^2+\mathbf{H}^2)dV$:
\begin{eqnarray}
{\cal H}={\frac{1}{2}}\sum_{n=1}^{N}\sum_{m=1}^{N}(\mathbf{{P}}_{n}\mathbf{{P}}_{m}+
{{\omega}^2}\mathbf{Q}_{n}\mathbf{Q}_{m})=\\=\sum_{n=1}^N{\cal H}_{nn}+\sum_{n\neq m}^{N}\sum_{m
\neq{n}}^{N}{\cal H}_{nm},
\end{eqnarray}
where
\begin{eqnarray}
\mathbf{Q}_{n}=\left( \frac{V}{4\pi{c^2}}\right)^{1/2}(\mathbf{a}_{\mathbf{k}}e^{i{\mathbf{k}
{\mathbf{r}+i\varphi_{n}}}}+\mathbf{a^*}_{\mathbf{k}}e^{-i{\mathbf{k}
{\mathbf{r}-i\varphi_{n}}}})
\end{eqnarray}
and
\begin{eqnarray}
\mathbf{P}_{n}=-i{\omega }\left( \frac{V}{4\pi{c^2}}\right)^{1/2}(\mathbf{a}_{\mathbf{k}}e^{i{\mathbf{k}
{\mathbf{r}+i\varphi_{n}}}}-\mathbf{a^*}_{\mathbf{k}}e^{-i{\mathbf{k}
{\mathbf{r}-i\varphi_{n}}}})
\end{eqnarray}
are the canonical variables, and $V$ is the volume. Notice that the energy ${\cal E}$ is found by integrating the energy flux (intensity) $\vec{S}$ of the field. To calculate
the Hamiltonian of spherical waves, in the potential
$\mathbf{A}_n(\mathbf{r},t)$, the term $e^{{\pm}i{\mathbf{k}} {\mathbf
{r}}\pm i\varphi_{n}}$ should be replaced by
$r^{-1}e^{{\pm}i{\mathbf{k}} {\mathbf{r}}\pm i{\varphi_{n}}}$. 

The first term in the expression (2) is the traditional Hamiltonian of
a classical single-mode EM field~\cite{Land,Loud,Bere,Cohe,Wein,Pike,Scul}. The
Hamiltonian does not take into account the interference phenomenon
(phase correlation) between the waves. The total energy of an EM field described by
the Hamiltonian is given by ${\langle \cal H \rangle}=\sum_{n=1}^N{\langle \cal
H \rangle}_{nn}=N{\cal E}_{1}$. Although the second term plays a key role in description of the enhancement and suppression of the light intensity in the interference phenomenon, the field theory does neglect the term under consideration of the field energy. For an example, in the case of a conventional Young's two-source setup, the interference cross-correlation term (energy) vanishes because of the fast spatial oscillations of the field in the transverse direction. The cross-correlation energy of an ensemble of the non-correlated (phase non-coherent) waves is also zero. In a general case, the field theory vanishes the cross-correlation energy artificially by considering the waves that satisfy the boundary conditions of an optical resonator~\cite{Land}. The resonator modes are orthogonal to each other. According to this approach, a light wave (beam) can be modulated with the superluminal velocity by the boundaries placed infinitely far from the beam. Nevertheless, the traditional models perfectly describe practically the all wave phenomena. In contrast to the traditional field theories we do not impose artificially the boundary conditions that are appropriate to optical resonators. In our model the light waves propagate in the free space as optical beams that diffracts and broaden. The positive or negative cross-correlation energy (second term) is responsible for the enhancement or suppression of energy
associated with the phase correlation between the waves. The energy can be increased or completely destroyed 
($0\leq{\langle \cal H \rangle}\leq N^2{\cal E}_{1}$) in an
ensemble of phase-coherent waves by modification of the wave phases $\varphi_{n}$. At appropriate
conditions, the phase modification may require an amount of energy
that is negligible compared to the energy of waves (see, Eq.~2). If the waves interfere 
destructively in all points of a physical system, the interference of waves completely destroys the
energy. The interference increases the energy if the waves interfere only constructively.

The quantum EM field is described by the quantized vector potential that has the form~\cite{Bere}:
\begin{eqnarray}
{\hat \mathbf{A}_n}=\hat
\mathbf{a}_{\mathbf{k}}{\mathbf{A}}_{\mathbf{k}}+{\hat
\mathbf{a}_{\mathbf{k}}}^{\dagger }{\mathbf{A}_{\mathbf{k}}}^*_{\mathbf{k}},
\end{eqnarray}
where
\begin{eqnarray}
{\mathbf{A}}_{\mathbf{k}}=\left( \frac{{2\pi{c^2}}}{\hbar \omega V}\right) ^{1/2}e^{i{\hbar \mathbf{k}{\mathbf{r}+i\varphi_{n}}}}.
\end{eqnarray}
The traditional Hamiltonian of the quantum field ${\hat
\mathbf{A}}$ is given by
\begin{eqnarray}
\hat {\cal H}={\frac{1}{8\pi }}{\int}({\hat \mathbf{E}}^2+{\hat
\mathbf{H}}^2)dV.
\end{eqnarray}
The quantum form of the Hamiltonian (7) can be found by using
the standard procedure~\cite{Land,Bere} based on the replacement of
the canonical variables (3) and (4) in the expression (1) by the operators
$\mathbf{\hat Q}_{n}$ and $\mathbf{\hat P}_{n}$:
\begin{eqnarray}
\hat {\cal H}={\frac{\hbar \omega}{2}}\sum_{n=1}^{N}\sum_{m=1}^{N}(\mathbf {\hat a^{\dagger}}_{\mathbf{k}}
e^{-i\varphi_{m}}\mathbf{\hat a}_{\mathbf{k}}
e^{i\varphi_{n}}+\mathbf{\hat a}_{\mathbf{k}}
e^{i\varphi_{m}}\mathbf{\hat a^{\dagger }}_{\mathbf{k}}
e^{-i\varphi_{n}}),
\end{eqnarray}
where $\mathbf{\hat a^{\dagger }}_{\mathbf{k}}$ and $\mathbf{\hat
a}_{\mathbf{k}}$ are the Dirac creation and destruction operators,
respectively. In order to take into account the interference phenomenon (the phase correlation between the waves) in both the field intensity and energy, we generalized the conventional commutation
relations: $[\mathbf{\hat a}_{\mathbf{k}n},\mathbf{\hat
a^{\dagger }}_{\mathbf{k}m}]$=1 and $[\mathbf{\hat
a}_{\mathbf{k}n},\mathbf{\hat a}_{\mathbf{k}m}]$=$[\mathbf{\hat
a^{\dagger }}_{\mathbf{k}n},\mathbf{\hat a^{\dagger }}_{\mathbf{k}m}]$=0.
The Hamiltonian (9) can be written also in the more convenient form:
\begin{eqnarray}
\hat {\cal H}=\sum_{n=1}^N\hat {\cal H}_{nn}+\sum_{n\neq m}^{N}\sum_{m\neq{n}}^{N}\hat {\cal H}_{nm},
\end{eqnarray}
where
\begin{eqnarray}
\hat {\cal H}_{nn}={\hbar \omega} \left( {\hat {\cal N}}_{\mathbf{k}}+{\frac{1}{2}} \right)
\end{eqnarray}
and
\begin{eqnarray}
\hat {\cal H}_{nm}={\frac{\hbar \omega}{2}}[\mathbf{\hat {\cal N}}_{\mathbf{k}}
e^{-i\varphi_{m}+i\varphi_{n}}+\nonumber\\
+({\hat {\cal N}}_{\mathbf{k}}+1)e^{i\varphi_{m}-i\varphi_{n}}].
\end{eqnarray}
Here, $\mathbf{\hat {\cal N}}_{\mathbf{k}}=\mathbf{\hat
a^{\dagger }}_{\mathbf{k}} \mathbf{\hat a}_{\mathbf{k}}$ is the photon
number operator. Notice that the vacuum-energy part in Eq.~(11) is a
real value under the alternative commutation relations:
$[\mathbf{\hat a}_{\mathbf{k}n},\mathbf{\hat
a^{\dagger }}_{\mathbf{k}m}]=\pm {e^{-i\varphi_{m}+i\varphi_{n}}}$ and
$[\mathbf{\hat a}_{\mathbf{k}n},\mathbf{\hat a}_{\mathbf{k}m}]$=$[\mathbf{\hat
a^{\dagger }}_{\mathbf{k}n},\mathbf{\hat a^{\dagger }}_{\mathbf{k}m}]$=0.
The two descriptions are different only in the vacuum energy.

In Eq.~9, the first term represents the traditional Hamiltonian of a
quantum single-mode EM field \cite{Land,Loud,Bere,Cohe,Wein,Pike,Scul}.
The Hamiltonian does not take into account the quantum interference between the different
waves. The total energy of an ensemble of the $N$ waves
describing by the traditional quantum Hamiltonian is given by $\langle {\cal E}
\rangle=N{\hbar \omega}(\langle{{\cal
N}}_{\mathbf{k}}\rangle+{\frac{1}{2}})$. The field contains $N\langle{{\cal N}}_{\mathbf{k}}\rangle$
particles, Einstein's photons having the energy $\hbar \omega$ and
momentum $\hbar \mathbf{k}$. The wave function of the field is constructed by using Fock's (number) phase-noncorrelated states $|n\rangle$. 
In the case of a quantum EM field describing by the full Hamiltonian (9), the second term is
responsible for the enhancement or suppression of the field intensity and energy associated with the
phase correlations between the waves. The interference provides the additional (from a point of view of the traditional models) the positive or negative cross-correlation energy. The energy can be increased or 
completely destroyed ($0\leq{\langle {\cal E} \rangle}\leq N^2{\hbar
\omega}(\langle{{\cal N}}_{\mathbf{k}}\rangle+{\frac{1}{2}})$) by interfering 
the $N$ quantum waves. The
respective number of photons could vary from zero to $N^2\langle{{\cal
N}}_{\mathbf{k}}\rangle$. According to our model, 
the field quantum state should be considered as a superposition of phase-correlated 
states (entangled state). Notice that $[\mathbf{\hat a}_{\mathbf{k}},\mathbf{\hat
a^{\dagger }}_{\mathbf{k}}]\approx 0$ for the big values of $\langle{{\cal
N}}_{\mathbf{k}}\rangle$. In such a case, the energy calculated by
the quantum Hamiltonians (9) tends towards the classical value (1).

So far we have considered a superposition of the single-mode
waves. To find the energy of a multi-mode classical or quantum
field one should use the above model for the multimode vector
potential having the form
$\mathbf{A}(\mathbf{r},t)=\sum_{\mathbf{k}}{\mathbf{A}_{\mathbf{k}}(\mathbf{r},t)}$,
where
${\mathbf{A}_{\mathbf{k}}(\mathbf{r},t)}=\sum_{n=1}^{N_{\mathbf{k}}}\mathbf{a}_{\mathbf{k}}e^{i{\mathbf{k}
{\mathbf{r}+i\varphi_{n\mathbf{k}}}}}+\mathbf{a^*}_{\mathbf{k}}e^{-i{\mathbf{k}
{\mathbf{r}-i\varphi_{n\mathbf{k}}}}}$. The summation is performed
over a set of values of the wave vector $\mathbf{k}$, where the values
are determined by the boundary conditions. For the sake of
simplicity, we present the quantum Hamiltonian for the
two-mode (${\mathbf{k_1}}\neq{\mathbf{k_2}}$) field $\mathbf{A}(\mathbf{r},t)= {\mathbf{A}}_{\mathbf
k_1}(\mathbf{r},t)+{\mathbf A}_{\mathbf k_2}(\mathbf{r},t)$:
\begin{eqnarray}
\hat {\cal H}={\hbar \omega}_1\left( {\hat {\cal N}}_{\mathbf{k_1}}+{\frac{1}{2}}\right) +
{\hbar \omega}_2\left( {\hat {\cal N}}_{\mathbf{k_2}}+{\frac{1}{2}}\right)+ \\
+{\frac{{\hbar} ( \omega_1 \omega_2)^{1/2}}{2}} {\left(
{\frac{\mathbf{\hat a^{\dagger }}_{\mathbf {k_1} } \mathbf {\hat
a}_{\mathbf {k_2}}}{{V}}}
{\int}e^{-i{\mathbf{k}_1{\mathbf{r}-i\varphi_{1}}}}
e^{i{\mathbf{k}_2{\mathbf{r}+i\varphi_{2}}}}dV \right) + } \\
+{\frac{{\hbar} (\omega_1 \omega_2)^{1/2}}{2}} {\left(
{\frac{\mathbf{\hat a}_{\mathbf {k_1} } \mathbf {\hat
a^{\dagger }}_{\mathbf {k_2}}}{{V}}}
{\int}e^{i{\mathbf{k}_1{\mathbf{r}+i\varphi_{1}}}}e^{-i{\mathbf{k}_2{\mathbf{r}-i\varphi_{2}}}}dV
\right)}+\\
+{\frac{{\hbar} ( \omega_1 \omega_2)^{1/2}}{2}} {\left(
{\frac{\mathbf{\hat a^{\dagger }}_{\mathbf {k_2} } \mathbf {\hat
a}_{\mathbf {k_1}}}{{V}}}
{\int}e^{-i{\mathbf{k}_2{\mathbf{r}-i\varphi_{2}}}}
e^{i{\mathbf{k}_1{\mathbf{r}+i\varphi_{1}}}}dV \right) + } \\
+{\frac{{\hbar} (\omega_1 \omega_2)^{1/2}}{2}} {\left(
{\frac{\mathbf{\hat a}_{\mathbf {k_2} } \mathbf {\hat
a^{\dagger }}_{\mathbf {k_1}}}{{V}}}
{\int}e^{i{\mathbf{k}_2{\mathbf{r}+i\varphi_{2}}}}e^{-i{\mathbf{k}_1{\mathbf{r}-i\varphi_{1}}}}dV
\right)}.
\end{eqnarray}
The term (12) is the traditional Hamiltonian of a quantum multimode EM field
\cite{Land,Loud,Bere,Cohe,Wein,Pike,Scul}. The term
is responsible for the interference of a quantum mode with itselve.
The cross-correlation term (13-16) is responsible for the intermode correlation
(interference) phenomenon. 
The cross-correlation integrals (13-16) could be
interpreted as exchange ones. The integrals describe the
quantum exchange of photons (interference) associated with the
indistinguishability of identical particles. The quantum exchange
of photons is somewhat similar to the exchange of virtual
particles for a short time ($\Delta t \leq 1/\Delta \cal E $) 
in perturbation theory. The cross-correlation integrals
have nonzero values if $\mathbf {k}_1 \approx \mathbf {k}_2$ or
$V=$ $\Delta x\Delta y\Delta z \leq 1 / (k_1-k_2)_x
(k_1-k_2)_y (k_1-k_2)_z$. Although the energy enhancement or suppression attributed to 
the cross-correlation integrals (13-16) is relevant to the energy uncertainty, 
the phenomenon should not be confused with the Heisenberg uncertainty 
principle. In quantum mechanics, lack of commutation of the time derivative operator with the time
operator itself mathematically results in an uncertainty principle
for time and energy: the longer the period of time, the more
precisely energy can be defined. Also notice that, in agreement with principle of the indistinguishability of
individual bosons, the commutation relations introduced above have
the canonical form if $\mathbf {k}_1 = \mathbf {k}_2$. It can be mentioned that the equations (1-16) can be easily rewrited for the
particular cases of non-coherent or partially coherent 
classical or quantum waves.

Let us now demonstrate that the interference-induced enhancement and 
suppression of the field energy do associate with several basic optical phenomena in a classical (non-quantum) EM field. A simple analysis of
Eqs.~(1-4) shows that the experimental realization of the phase
conditions required for the interference-induced enhancement or 
suppression of the field energy is practically impossible in 
conventional (non-subwavelength) optical systems (also, see 
the study \cite{Nott} and references therein). In the study
\cite{Kukh}, however, we have showed that the waves generated by
the point-like sources separated by the distance
${\Lambda}<\lambda $ satisfy
the phase conditions in the far-field diffraction zone. One can now 
easily demonstrate by using Eqs.~(1-4) that the
interference-induced enhancement or suppression of the field energy could be 
associated with the extraordinary transmission
of light through subwavelength apertures in metal screens
\cite{Ebb,Barn,Schr,Sobn,Port,Asti,Taka,Lala,Barb,Kukh1,Leze,More,Liu,Garc,Tre,Cao,Sar,
Bar,Koer,Miro,Pend,Gome,Mo,Ung,Hua,Ben,Sary}. Indeed, in such a kind of experiments, the wave vectors of light
waves produced by the subwavelength apertures are practically the
same, $\mathbf k_n\approx\mathbf k_1$. In the case of the subwavelength-dimension 
apertures separated by the distance
${\Lambda}<<\lambda $, the cross correlation term in
Eq. (1) enhances the field energy. Another example is  
the enhancement or suppression of the field energy of a light pulse
(wavepacket) propagating in a dispersive medium. The phases of the
different Fourier $\mathbf k$-components (the components propagate in the same
direction, $\mathbf k_n/k_n=\mathbf k_1/k_1$) of the wavepacket can be changed
under the propagation. According to Eqs. (1-4) the phase
modification could lead to the interference-induced enhancement or suppression of 
the pulse energy. The energy can be enhanced or completely destroyed if the amount of energy spent on the
phase modification is smaller than the energy of the Fourier
components. The interference phenomenon could be relevant to the variation of
the momentum of light in a material medium
\cite{Leo} and the energy variation in an ultra-short light pulse scattering by
subwavelength apertures \cite{Kuk}. Taking into account the cross
correlation energy could be important also for understanding the enhancement and suppression of energy
in other optical processes (for example, see the studies~\cite{Nott,Gau,Gord,Sch} and references therein). 
We stress again that the cross-correlation energy vanishes if the light waves satisfy 
the boundary conditions of a cavity resonator. The waves (resonator modes) having the
spatial distributions appropriate for an optical resonator are orthogonal to each other.

The interference-mediated enhancement and suppression of the field energy
do associate with several basic quantum
phenomena also in a quantum EM field. As an example, let us consider the energy of two photons. It is
generally accepted that the energy of two photons is always given by 
${\langle {\cal E} \rangle}= 2{\hbar \omega}$ (see, the first term of the Hamiltonian (12)). A simple analysis of the full Hamiltonian (12-16) shows that
the energy of two phase-correlated coherent photons (a biphoton)
with ${\hbar \omega}_1={\hbar \omega}_2={\hbar \omega}$ depends on the wave
vectors and phases, $0\leq{\langle {\cal E} \rangle}\leq 4{\hbar
\omega}$. According to our model the energy of two non-corrlelated photons only is
given by ${\langle {\cal E} \rangle}= 2{\hbar \omega}$. The fact
that energy (wavelength $\lambda =2\pi \hbar c/{\langle {\cal E} \rangle}$)
of a biphoton depends on the experimental
configuration of the biphoton generation, in our model on the wave
vectors and phases,
is well known~\cite{So}. In Ref. \cite{Kukh}, we have predicted
the enhancement or suppression of the field energy by any subwavelength-dimension
physical system taking into account the interference properties
of Young's double-source subwavelength system. It is important to
consider the interference properties by using the full quantum Hamiltonian
(12-16). In conventional Young's setup, the two plane waves
generated by the two pinholes separated by the distance
${\Lambda}>>\lambda$ have different wave vectors, ${\mathbf
k_1}\neq {\mathbf k_2}$ \cite{Kukh}. According to the expressions (13-16),
the cross correlation term vanishes. Respectively, the
field energy is given by ${\langle {\cal E} \rangle}= 2{\hbar \omega}$. In the case of Young's
subwavelength system ${\Lambda}<<\lambda$, correspondingly ${\mathbf
k_1}={\mathbf k_2}$. Thus, in the case of $\varphi_1-\varphi_2=0$, the
interference creates the extra energy, ${\cal E}=4{\hbar\omega}$. 
The interference completely destroys
the energy at the phase condition $\varphi_1-\varphi_2=\pi$. 
If the spacing of the apertures is larger (${\Lambda}>\lambda$), then the addition of waves is not so efficient, but still leads to enhancements and resonances (versus wavelength) in the total power transmitted.
In such a case the biphoton energy ($0\leq{\langle {\cal
E} \rangle}\leq 4{\hbar \omega}$) depends on the values $\varphi_1$, $\varphi_2$ and ${\Lambda}$. 
A simple analysis of the Hamiltonian
(12-16) also shows that a quantum entangled (phase-correlated) state of photons is
preserved on passage through an aperture array. The
propagation of an entangled) state through an optically
lens, however, destroys the correlation by the well-known
modification of the wave vectors ${\mathbf k}$ and phases
$\varphi_{\mathbf k}$. Such a behavior is in agreement with the
recent experiment \cite{Alte}. Finally, let us consider a Dicke
superradiance quantum model \cite{Dick} of emission of a
subwavelength ensemble of atoms. The wave vectors of
the light waves produced by the subwavelength-dimension ensemble of 
atoms in the far-field zone are practically the same, $\mathbf k_n\approx\mathbf k_1$. 
According to Ref.~\cite{Dick} and the Hamiltonian (12-16) 
the light energy scales as the number of light-sources squared, 
regardless of periodicity ($\langle {\cal E} \rangle=N^2{\hbar \omega}$, where $N$ is
the number of the phase-correlated atoms or photons). In addition to 
the superradiance, our model predicts destruction of the field energy at 
the phase condition $\varphi_1-\varphi_2=\pi$ for the atom or photon pairs. 
At such a phase condition the original photon pairs in the quantum 
field have to be vanished. A simple analysis of Eqs. 12-16 also shows that the interference of quantum waves at the detector in the case of ${\Lambda}>\lambda$ could lead to a new effect, namely the enhancements and resonances (versus period of the array) in the total power emitted by the periodic array of quantum oscillators (atoms). In the present study, we have considered photon fields. A quantum reformulation of our model for other boson and fermion fields will be presented in our next paper.

In conclusion, the Hamiltonian describing the interference-induced enhancement or suppression of the intensity and energy of a multimode quantum optical field was derived. Such a field can be produced, for instance, by a subwavelength array of coherent light sources. We have
showed that the interference phonomenon do associate with many
basic optical phenomena, such as the extraordinary transmission
of light through subwavelength apertures, scattering of entangled
photons in Young's two-slit experiment and Dicke's quantum
superradiance.

This study was supported in part by the Framework for European Cooperation in the field of Scientific and Technical Research (COST, Contract No MP0601) and the Hungarian Research and Development Program (KPI, Contract GVOP 0066-3.2.1.-2004-04-0166/3.0).

\end{document}